# Enhanced Transmission and Second Harmonic Generation from Subwavelength Slits on Metal Substrates

M.A. Vincenti[a,b], M. De Sario[a], V. Petruzzelli[a], A. D'Orazio[a], F. Prudenzano[c], D. de Ceglia[b], N. Akozbek[b], M.J. Bloemer[b], P. Ashley[b], M. Scalora[b]

[a]Dipartimento di Elettrotecnica ed Elettronica, Politecnico di Bari, Via Orabona 4, 70125 Bari, Italy
[b]Charles M. Bowden Research Center, AMSRD-AMR-WS-ST, RDECOM, Redstone Arsenal, Alabama 35898-5000, USA
[c]Dipartimento di Ingegneria dell'Ambiente e per lo Sviluppo Sostenibile, Politecnico di Bari, Viale del Turismo 8, Taranto, Italy

**ABSTRACT**

We theoretically investigate second harmonic generation that originates from the nonlinear, magnetic Lorentz force term from single and multiple apertures carved on thick, opaque metal substrates. The linear transmission properties of apertures on metal substrates have been previously studied in the context of the extraordinary transmission of light. The transmission process is driven by a number of physical mechanisms, whose characteristics and relative importance depend on the thickness of the metallic substrate, slit size, and slit separation. In this work we show that a combination of cavity effects and surface plasmon generation gives rise to enhanced second harmonic generation in the regime of extraordinary transmittance of the pump field. We have studied both forward and backward second harmonic generation conversion efficiencies as functions of the geometrical parameters, and how they relate to pump transmission efficiency. The resonance phenomenon is evident in the generated second harmonic signal, as conversion efficiency depends on the duration of incident pump pulse, and hence its bandwidth. Our results show that the excitation of tightly confined modes as well as the combination of enhanced transmission and nonlinear processes can lead to several potential new applications such as photo-lithography, scanning microscopy, and high-density optical data storage devices.

**Keywords:** Enhanced Transmission, Second Harmonic Generation, Surface Plasmon, Sub-wavelength Slit

## 1. INTRODUCTION

Extraordinary transmission and light scattering through one or more subwavelength holes in thick opaque metals have been extensively studied in order to understand transmission and diffraction mechanisms from the optical to the microwave regime [1-5]. Ebbesen *et al* showed that extraordinary transmission through an array of sub-wavelength holes can occur under a particular condition [6] and that this phenomenon was based exclusively on plasmon excitation. However, since the original discovery, this field has witnessed some degree of controversy, because several effects can contribute to this phenomenon. For instance, it has been discovered that both TE and TM modes actually experience the extraordinary transmission process, and that cavity effects (the hole becomes a resonant cavity) contribute for both polarizations states [3]. Several theoretical studies have reported on the transmission and reflection of light through subwavelength holes on *perfect electric conductor* (PEC) screens [4-7]. They showed that the calculation of the propagation constant and the wavelength of the Fabry-Perot resonance inside the slit are easily done. However, as predicted by surface plasmon theory [8], the introduction of a predominant negative real part of the relative permittivity of the metal results in a shift of the well known Fabry-Perot resonances [4, 9, 10]. Linear transmission processes basically depend on a number of physical mechanisms, such as the thickness of the metallic substrate, slit size, and slit separation. In this regime of extraordinary transmission of the pump field we can exploit a combination of cavity effects and longitudinal surface plasmon generation leading to enhanced second harmonic generation. Even if metals do not have intrinsic quadratic nonlinear terms, several works report on a second harmonic generation in metals based on the presence of the Lorentz force induced by the magnetic field: in fact, using a classical oscillator electron model, a second harmonic source term consisting of a magnetic dipole due to the Lorentz force and of an electric quadrupole contribution due to the Coulomb force can be observed also in centrosymmetric media [11, 12]. Subsequent experimental work confirmed the existence of two second harmonic source terms [13, 14], a volume and a surface source term that can be excited respectively by a polarization normal or parallel to the plane of incidence. However, in the case of bulk metals one can consider the surface effect as being dominant and responsible for most of the generated signal. Second harmonic generation can be significantly enhanced by coupling with surface plasmons: second harmonic generation by reflection of the incident beam from a metal surface has important contributions from currents stimulated in the immediate vicinity of the surface. The current confined at the metal surface has two components, one normal and one tangential to the surface, which extends only a few Fermi wavelengths inside the metal. In a recent theoretical and experimental study of second harmonic generation (SHG) an enhancement of second harmonic generation by at least a factor of 30 was demonstrated compared to SHG by a single metal layer from a specifically designed metal-dielectric photonic band gap structure [15]. The results of reference [15] suggest that field localization properties make it possible for volume contribution to become of the same order as surface contributions. These results have the potential of leading to several new applications such as photo-lithography, scanning microscopy, high-density optical data storage devices or efficient biological and chemical sensors.

## 2. ENHANCED TRANSMISSION AND SECOND HARMONIC GENERATION FROM A SINGLE SLIT

The transmission process in a single subwavelength slit carved on a thick metal substrate is driven by two different mechanisms, i.e. the thickness of the metallic substrate and the slit size. The structure under investigation is depicted in Fig.1: it consists of a silver layer whose dispersion profile is found in Palik's handbook [16], and whose parameters under investigation are $w$ and $a$.

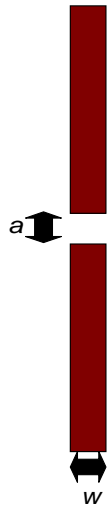

**Fig.1:** Sketch of the silver substrate: the parameters under investigation are $a$ (slit size) and $w$ (substrate thickness)

The variation of these two parameters is examined in order to enhance the linear transmittance and to begin to appreciate the interaction between the geometrical and physical mechanisms. All the calculations and results reported below are obtained solving Maxwell's equations by two independent means: (i) a home-made finite-difference, time-domain (FDTD) method, whose details are discussed in reference [17]; and (ii) a time-domain fast Fourier transform beam propagation (FFT-BPM) method, which takes a completely different approach, and yet yields nearly identical results as the FDTD [15, 18, 19]. Since the FDTD method is the most widely adopted technique, we now outline the FFT-BPM method because it is a potent alternative with great flexibility [15].

In Gaussian units, the system of equations we aim to solve is as follows:

$$\nabla \times \mathbf{E} = -\frac{1}{c}\frac{\partial \mathbf{B}}{\partial t} \qquad \nabla \times \mathbf{H} = \frac{1}{c}\frac{\partial \mathbf{E}}{\partial t} + \frac{4\pi}{c}\frac{\partial \mathbf{P}}{\partial t}$$
$$\ddot{\mathbf{P}} + \gamma \dot{\mathbf{P}} = \frac{\omega_p^2}{4\pi}\mathbf{E} + \frac{e}{mc}\dot{\mathbf{P}} \times \mathbf{H}$$
(1)

where the last term in the polarization equation is the nonlinear, magnetic component of the Lorentz force. We assume a right-handed coordinate system, and $\hat{p}$-polarized (TM) pump and second harmonic fields of the type:

$$\mathbf{E} = \begin{pmatrix} \mathbf{j}\left(E_y^\omega e^{-i\omega t} + \left(E_y^\omega\right)^* e^{i\omega t} + E_y^{2\omega} e^{-2i\omega t} + \left(E_y^{2\omega}\right)^* e^{2i\omega t}\right) + \\ \mathbf{k}\left(E_z^\omega e^{-i\omega t} + \left(E_z^\omega\right)^* e^{i\omega t} + E_z^{2\omega} e^{-2i\omega t} + \left(E_z^{2\omega}\right)^* e^{2i\omega t}\right) \end{pmatrix}$$

$$\mathbf{H} = \mathbf{i}\left(H_x^\omega e^{-i\omega t} + \left(H_x^\omega\right)^* e^{i\omega t} + H_x^{2\omega} e^{-2i\omega t} + \left(H_x^{2\omega}\right)^* e^{2i\omega t}\right)$$
(2)

The corresponding macroscopic polarization is given by:

$$\mathbf{P} = (P_y \mathbf{j} + P_z \mathbf{k}) = \begin{pmatrix} \mathbf{j}\left(P_y^\omega e^{-i\omega t} + \left(P_y^\omega\right)^* e^{i\omega t} + (P_y^{2\omega} e^{-2i\omega t} + \left(P_y^{2\omega}\right)^* e^{2i\omega t}\right) + \\ \mathbf{k}\left(P_z^\omega e^{-i\omega t} + \left(P_z^\omega\right)^* e^{i\omega t} + P_z^{2\omega} e^{-2i\omega t} + \left(P_z^{2\omega}\right)^* e^{2i\omega t}\right) \end{pmatrix}$$
(3).

The envelope functions $E_y^{\omega,2\omega}$, $E_z^{\omega,2\omega}$, $H_x^{\omega,2\omega}$, $P_y^{\omega,2\omega}$, $P_z^{\omega,2\omega}$ contain implicit spatial dependences that for simplicity have been omitted. In addition, the envelope functions are not assumed to be slowly varying, as no approximations are made when the fields and polarizations are substituted into Maxwell's equations. In other words, the decomposition of the fields in Eqs.(2-3) is a matter of mere convenience. For second harmonic generation, substitution of Eqs.(2-3) into Eqs.(1) results in a system of fourteen coupled differential equations for the fields, polarizations, and corresponding currents, which in scaled form are written as follows for the pump:

$$\frac{\partial H_x^\omega}{\partial \tau} = i\beta\left(H_x^\omega + E_z^\omega \sin\theta_i + E_y^\omega \cos\theta_i\right) - \frac{\partial E_z^\omega}{\partial \tilde{y}} + \frac{\partial E_y^\omega}{\partial \xi}$$

$$\frac{\partial E_y^\omega}{\partial \tau} = i\beta\left(E_y^\omega + H_x^\omega \cos\theta_i\right) + \frac{\partial H_x^\omega}{\partial \xi} - 4\pi(J_y^\omega - i\beta P_y^\omega)$$

$$\frac{\partial E_z^\omega}{\partial \tau} = i\beta\left(E_z^\omega + H_x^\omega \sin\theta_i\right) - \frac{\partial H_x^\omega}{\partial \tilde{y}} - 4\pi(J_z^\omega - i\beta P_z^\omega)$$

$$\frac{\partial J_y^\omega}{\partial \tau} = (2i\beta - \gamma_\omega)J_y^\omega + (\beta^2 + i\gamma_\omega\beta)P_y^\omega + \frac{\pi\omega_{p,\omega}^2}{\omega_r^2}E_y^\omega$$
$$+ \frac{e\lambda_0}{mc^2}\left[\left((J_z^\omega)^* + i\beta(P_z^\omega)^*\right)H_x^{2\omega} + \left(J_z^{2\omega} - 2i\beta P_z^{2\omega}\right)(H_x^\omega)^*\right]$$

$$\frac{\partial J_z^\omega}{\partial \tau} = (2i\beta - \gamma_\omega)J_z^\omega + (\beta^2 + i\gamma_\omega\beta)P_z^\omega + \frac{\pi\omega_{p,\omega}^2}{\omega_r^2}E_z^\omega$$
$$- \frac{e\lambda_0}{mc^2}\left[\left((J_y^\omega)^* + i\beta(P_y^\omega)^*\right)H_x^{2\omega} + \left(J_y^{2\omega} - 2i\beta P_y^{2\omega}\right)(H_x^\omega)^*\right]$$

$$J_y^\omega = \frac{\partial P_y^\omega}{\partial \tau} \qquad J_z^\omega = \frac{\partial P_z^\omega}{\partial \tau} \qquad , \quad (4)$$

and as follows for the SH:

$$\frac{\partial H_x^{2\omega}}{\partial \tau} = 2i\beta\left(H_x^{2\omega} + E_z^{2\omega}\sin\theta_i + E_y^{2\omega}\cos\theta_i\right) - \frac{\partial E_z^{2\omega}}{\partial \tilde{y}} + \frac{\partial E_y^{2\omega}}{\partial \xi}$$

$$\frac{\partial E_y^{2\omega}}{\partial \tau} = 2i\beta\left(E_y^{2\omega} + H_x^{2\omega}\cos\theta_i\right) + \frac{\partial H_x^{2\omega}}{\partial \xi} - 4\pi(J_y^{2\omega} - 2i\beta P_y^{2\omega})$$

$$\frac{\partial E_z^{2\omega}}{\partial \tau} = 2i\beta\left(E_z^{2\omega} + H_x^{2\omega}\sin\theta_i\right) - \frac{\partial H_x^{2\omega}}{\partial \tilde{y}} - 4\pi(J_z^{2\omega} - 2i\beta P_z^{2\omega})$$

$$\frac{\partial J_y^{2\omega}}{\partial \tau} = (4i\beta - \gamma_{2\omega})J_y^{2\omega} + (4\beta^2 + i\gamma_{2\omega}2\beta)P_y^{2\omega} + \frac{\pi\omega_{p,2\omega}^2}{\omega_r^2}E_y^{2\omega} + \frac{e\lambda_0}{mc^2}\left(J_z^\omega - i\beta P_z^\omega\right)H_x^\omega$$

$$\frac{\partial J_z^{2\omega}}{\partial \tau} = (4i\beta - \gamma_{2\omega})J_z^{2\omega} + (4\beta^2 + i\gamma_{2\omega}2\beta)P_z^{2\omega} + \frac{\pi\omega_{p,2\omega}^2}{\omega_0^2}E_z^{2\omega} - \frac{e\lambda_0}{mc^2}\left(J_y^\omega - i\beta P_y^\omega\right)H_x^\omega$$

$$J_y^{2\omega} = \frac{\partial P_y^{2\omega}}{\partial \tau} \qquad J_z^{2\omega} = \frac{\partial P_z^{2\omega}}{\partial \tau}$$

. (5)

We note the lack of any approximations. We have chosen $\lambda_r = 1\mu m$ as the reference wavelength, and have adopted the following scaling: $\xi = z/\lambda_r$ and $\tilde{y} = y/\lambda_r$ are the scaled longitudinal and transverse coordinates, respectively; $\tau = ct/\lambda_r$ is the time in units of the optical cycle; $\beta = 2\pi\tilde{\omega}$ is the scaled wave vector; $\tilde{\omega} = \omega/\omega_r$ is the scaled frequency, and $\omega_r = 2\pi c/\lambda_r$, where $c$ is the speed of light in vacuum. $\theta_i$ is the angle of incidence of the pump with respect to the normal direction. The magnitude of the coupling coefficient in the Lorentz force term is evaluated in Gaussian units: $\frac{e\lambda_0}{mc^2} = \frac{(-4.8\times10^{-10})\times(10^{-4})}{(9.1\times10^{-28})\times(9\times10^{20})} = -5.9259\times10^{-8}(cgs)$. The linear dielectric response of silver has been assumed to be Drude-like, as follows: $\varepsilon(\tilde{\omega}) = 1 - \frac{\omega_P^2}{\tilde{\omega}^2 + i\gamma\tilde{\omega}}$. At 800nm, the actual data [16] is fitted using the set of parameters: $(\gamma_\omega, \omega_{p,\omega})$=(0.06, 6.73), and at 400nm we have: $(\gamma_{2\omega}, \omega_{p,2\omega})$=(0.33, 5.51). The incident magnetic field was assumed to be Gaussian of the form: $\mathcal{H}_x(\tilde{y}, \xi, \tau = 0) = H_0 e^{-[(\xi-\xi_0)^2 + \tilde{y}^2]/w^2}$, with similar expressions for the transverse and longitudinal electric fields. The equations are then solved in the time domain using the well-known, widely-used split-step algorithm. In short, the propagation is done in two separate steps: first the material interaction is taken into account, and then the fields are propagated in free space. The equations are integrated using a mix of a predictor-corrector algorithm for the material part, and fast Fourier transform for the free space propagation part. The main advantages that the FFT approach offers over the FDTD method may be summarized as follows: (1) calculation of the spatial derivative is much more accurate; (2) convergence is somewhat more rapid, in part due to the accurate rendition of the spatial derivatives, thus allowing for larger temporal and spatial integration steps; (3) the method is unconditionally stable, and allows one to choose a temporal integration step based on numerical convergence rather than longitudinal or transverse discretization steps.

For both numerical methods we considered a Gaussian-shaped incident pump tuned at 800nm, and having a peak intensity $I_{in}$=2GW/cm$^2$. By considering such source input, and by normalizing the transmitted field to the energy that impinges in the geometrical area of the slit, as is customarily done, we obtain a two-dimensional map of the transmittance that we depict in Fig.2. This calculation was performed by varying the parameter *a* between 32nm (*a* << λ/2) and 496nm (*a* > λ/2), while the thickness of the metallic substrate *w* varied between 88nm (*w* << λ/2) and 640nm (*w* > λ/2). As pointed out in Fig.2 the Fabry-Perot-like behavior of such structures is altered with respect to previously reported transmittance peculiarities of an equivalent PEC structure [7]. In fact, we find that for a slit size equal to ~32nm we can improve the linear transmission up to 3 times the linear transmission of the equivalent PEC substrate. This means that the presence of a longitudinal plasmon resonance inside the nanocavity also plays an equally important role in the resonant mechanism. In other words, the enhancement of the transmitted field is due to coupling that depends on two factors: (i) substrate thickness; and (ii) aperture size. Central to all this is the treatment of the metal as a realistic material, having finite dielectric constant and being penetrable by the wave. Fig.2 highlights several differences between what occurs for a PEC and a realistic material, i.e silver at optical wavelengths: small apertures carved on a perfect metal exhibit the same transmission maxima regardless of slit size, while for a realistic material Fabry-Perot resonances acquire a different qualitative texture. The figure shows that the transmission curve is quite complex and features several irregularly shaped maxima. For instance, an aperture size $a\sim$32nm displays an absolute maximum for $w\sim$200nm, and a relative maximum for $w\sim$440nm. Increasing aperture size shifts continuously the resonant conditions, which for larger apertures occurs at larger substrate thicknesses. For example, for $a\sim$64nm transmission maxima develop at $w\sim$215nm and $w\sim$540nm. Furthermore, Fig.2 confirms the influence of aperture size on the enhanced transmission phenomenon. In contrast, in a perfect metal this parameter does not alter the transmission process.

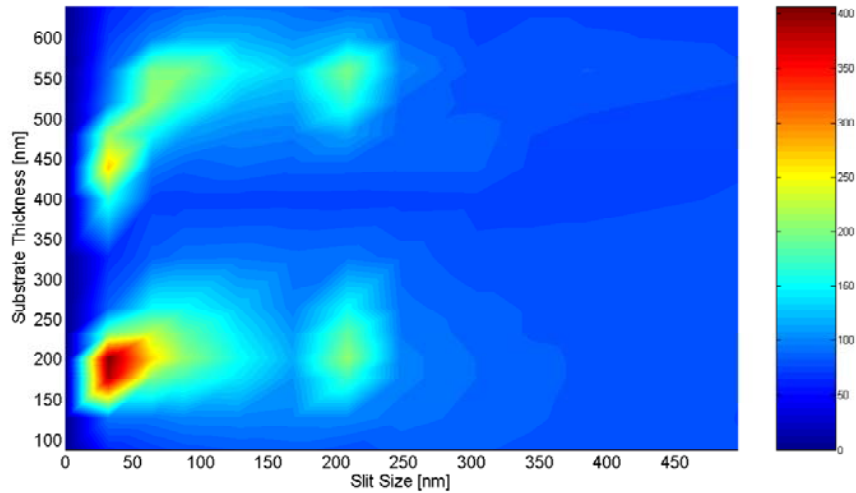

**Fig.2:** Two-dimensional map of the transmission coefficient evaluated by varying the substrate thickness and the aperture size. The presence of two resonances for the same substrate thickness confirms the Fabry-Perot like behavior of this structure.

It is well known that metals do not have intrinsic quadratic nonlinear term. However, several theoretical and experimental works report on the possibility to generate a second harmonic signal impinging on a metallic screen. This is a direct consequence of the presence of a magnetic dipole due to the Lorentz force and of an electric quadrupole contribution due to the Coulomb force [11, 12], as we outlined in the model presented above. It has been also demonstrated that the second harmonic generation can be further improved by varying the incident angle, finding an optimal condition for the generated signal [13, 14, 20]. In a recent work, it was shown that oit is possible to improve second harmonic generation by at least a factor of 30 compared to SHG from a single metal layer by means of metal-dielectric photonic band gap structure [15].

Our result suggests that the extraordinary transmittance regime is closely correlated with the transmitted SHG process. This is a strong indication that cavity effects are simultaneously important in linear and nonlinear processes that depend on the longitudinal localization of light. We note that if the metal under considerations is a realistic metal, then these enhancement effects occur without surface corrugations or multiple apertures. We begin our considerations by examining transmitted and reflected second harmonic signals. We place a single 32nm slit on a Ag substrate. As shown on Fig.3, we also monitor reflection from the slit. Linear pump transmission and transmitted SHG are easily correlated: pump transmittance is maximized when transmitted SHG is a maximum. At those wavelengths, the incident field couples rather well to the nanocavity, as field localization effects amplify the internal fields by factors of ~50. The simultaneity of field localization and plasmonic effects then combine to give ~410% extraordinary transmittance, and enhanced SHG with respect to the bare metal [15]. We also note that maximum linear reflectance and maximum linear transmittance do not coincide, as illustrated on Fig.3.

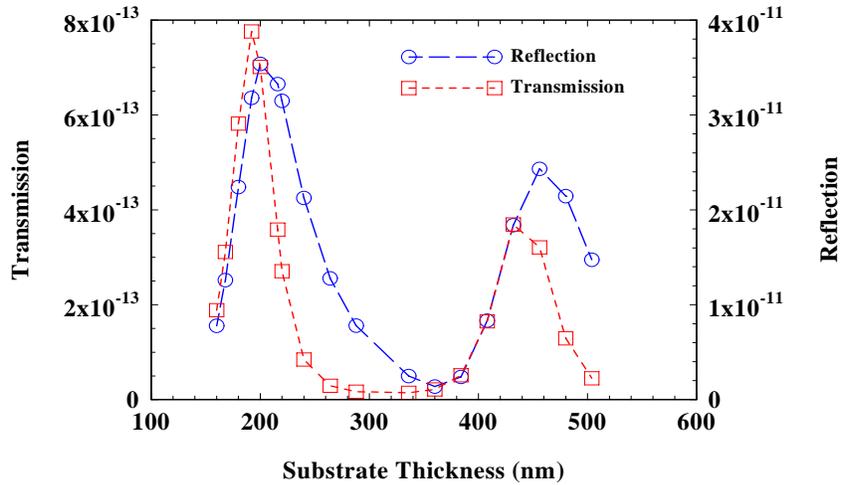

**Fig.3:** Second harmonic conversion efficiency for a single slit of 32nm carved on a silver substrate: the reflected signal (right axis) is approximately 50 times larger than the transmitted signal (left axis). We also note that maximum transmission and maximum reflection do not coincide: the transmission process is driven in part by a longitudinal resonance process, and in part by longitudinal surface plasmons that lead to extraordinary transmittance.

Field localization properties inside the 32nm-wide nanocavity for the FF and SH magnetic fields are reported in Figs.4a and 4b, respectively: a substrate thickness $w$~200nm maximizes and enhances linear transmittance up to nearly 410%, assures good coupling of the pump field inside the nanocavity, and yields enhanced transmitted SHG. This strong, forward-coupling condition for both the pump and the generated signal is progressively spoiled by modulating substrate thickness. Although the amount of SH reflection is nearly two orders of magnitude stronger than any transmitted SHG, it is also possible to modulate SH reflections by changing substrate thickness. The figure suggests that favorable conditions for reflected SHG are found when $w$~220nm.

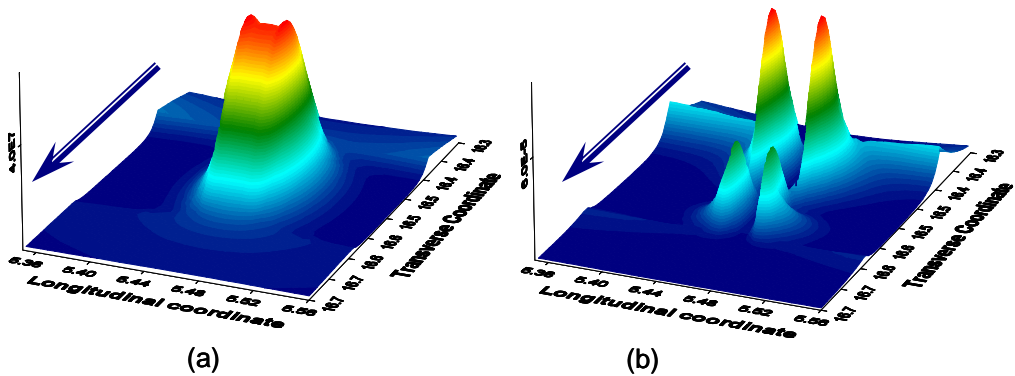

(a)  (b)

**Fig.4:** Field localization inside the nanocavity for a) the pump and b) the second harmonic signal. The arrows indicate the direction of propagation.

## 3. ENHANCED TRANSMISSION AND SECOND HARMONIC GENERATION FROM MULTIPLE APERTURES

In the last decades several works have shown that periodic metallic structures, such as array of slits [21] or single slit surrounded by periodic corrugation [22,23], also provide for transmission enhancement. It has been also demonstrated that periodical and random geometries are also useful to improve the nonlinear response of these structures [6, 24]. For both these reasons we investigated on the possibility to improve our transmission enhancement by using one-dimensional multiple aperture systems. The linear analysis is also followed by a thorough analysis of the nonlinear response of these geometries.

### 3.1 Two Apertures

In the previous section we showed a two-dimensional map that relates the slit size and the substrate thickness to the linear transmission peculiarities (Fig.2). Our calculations demonstrated that for a realistic metal such as silver, the Fabry-Perot resonances are not necessarily preserved at their expected positions when aperture size is changed. If we now consider a simple extension to a double-aperture system, one soon realizes that a third resonant mechanism emerges from this new geometry: the distance between the apertures. While surface waves exiting from the single aperture do not encounter obstacles upon propagating on the metal surfaces, it is clear that the double aperture system also involves this aspect. It has been theoretically [25] and experimentally [26] demonstrated that the metallic region between the slits plays a significant role for the transmission process. The process is easy to understand if we think of the space between apertures as resonant cavity for the surface waves. Once again, perfect metal conditions [25] lend themselves to a straight-forward theoretical treatment that is able to recover maxima and minima that depend on the distance between slits. In contrast, theoretical predictions of resonant behavior become much more difficult if the actual permittivity of the metal is introduced.

For we consider a silver substrate with two apertures, and whose key parameters (see Fig.5) are substrate thickness $w$, and center-to-center slit distance $s$. For convenience, we fix aperture size to maximize linear transmittance, i.e. $a$=32nm. All the calculations and results reported for the double aperture system are obtained by means of the same numerical methods mentioned above, and by considering the same Gaussian shaped incident pump field tuned at 800nm, and having a peak intensity $I_{in}$=2GW/cm$^2$.

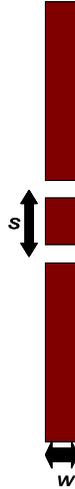

**Fig.5:** Sketch of the silver substrate with two apertures: while the aperture size is fixed at *a*=32nm, the key parameters are *w* (substrate thickness) and the separation between the slits *s* (center to center).

Just as was the case for the single slit system, the double aperture system is also influenced by the substrate thickness *w*: an optimal condition for the linear properties is obtained when $w \sim \lambda/4$ (Fig.6). This value of *w* allows further enhancement of the linear transmission from $T_{1slit} \sim 410\%$ to $T_{2slit} \sim 650\%$, reaching its maximum value when the slit separation is $s \sim \lambda/2$. Moreover, we note than the resonant peak shifts to smaller slit separations as the thickness of the metal substrate is increased. This means that constructive interference between the plasmons exiting from the two apertures changes dramatically for slight variations of *w*.

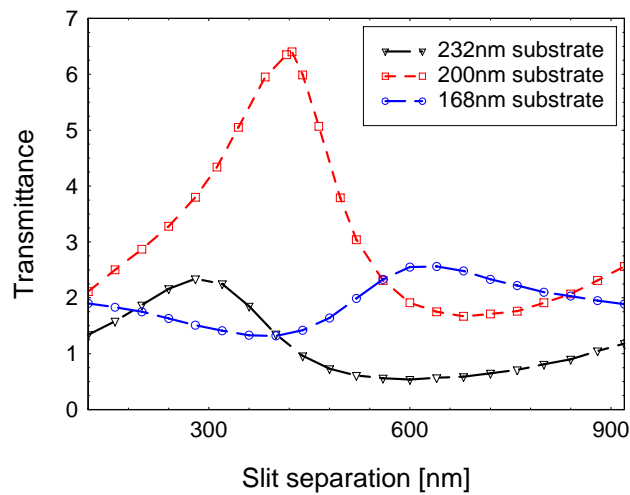

**Fig.6:** Transmittance versus separation between two 32nm-wide slits. By increasing the substrate thickness the resonance is shifted towards smaller values of *s*.

Here too we find that the nonlinear response is influenced by the linear transmission properties. As depicted on Fig.7, we find that the best conditions for the transmitted second harmonic signal occur when the linear response reaches its maximum. On the other hand, maximum reflected second harmonic generation is obtained when slit separation is increased considerably, when the plasmons inside the nanocavities are totally decoupled (transmittance is a minimum) and the interference between the plasmons exiting from the aperture is destructive. The conversion efficiency thus benefits from the enhancement of the pump resulting in an improvement of a factor of 10 for the transmitted signal, while the reflected signal is approximately doubled in its efficiency: this difference arises from the fact that the transmitted field depends also on the coupling effect between the apertures and not only on the plasmon coupling inside the nanocavities.

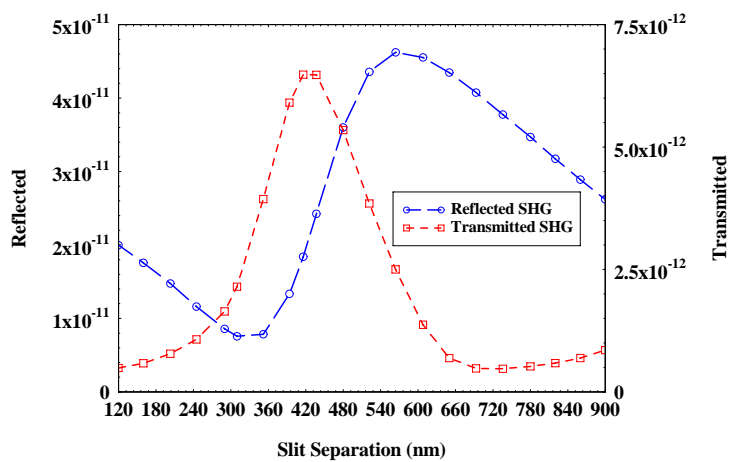

**Fig.7:** Reflected (left axis) and transmitted (right axis) second harmonic conversion efficiency for two 32nm slits carved on a 200nm silver substrate: the transmitted second harmonic signal benefits from the enhancement of the pump inside the cavity, resulting in an improvement of a factor of 10 in conversion efficiency relative to the single slit case.

### 3.2 Multiple Apertures

The results for the two aperture systems are indicative of what occurs when more apertures are added, in the sense that curves similar to Fig.7 are obtained for 4-, 8-, and 16-aperture systems. In Fig.8 we plot the linear transmittance of the system as a function of the number of apertures, and we see that saturation effects quickly take hold for very few apertures. Given the same inter-slit distance, the maximum enhanced transmission is just above 700%.

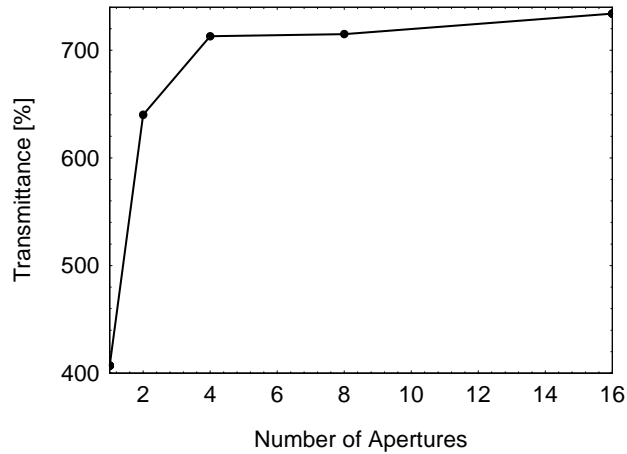

**Fig.8:** Comparison of the linear response against the number of apertures: the enhancement of the transmitted field is not significant increasing the number of the apertures

## 4. CONCLUSIONS

In this paper we have theoretically investigated the intriguing phenomenon of enhanced transmission that occurs for subwavelength apertures carved on metal substrates. The combination of an improved linear response together with the Lorentz force contribution leads to the generation and enhancement of the transmitted and reflected second harmonic signals that display resonance cavity effects. The geometrical parameters that we considered in our analysis were center-to-center distance between the slits, metal substrate separation, and number of slits. Our results suggest that enhanced linear transmission is obtained even for a single slit on a smooth metal surface, and that inside the aperture the fields may be enhanced by approximately a factor of 50. While we find that transmitted SHG is directly correlated to the enhancement of linear transmission, the reflected SH signal has somewhat different sensitivities, as scattering for the aperture and the smooth surface change its resonance conditions.